\documentclass[final,3p,12pt,times,sort&compress]{elsarticle}

\usepackage{color}
\usepackage{graphicx}
\usepackage{amsmath, mathtools}
\usepackage{amssymb}
\usepackage{graphicx}
\usepackage{grffile}
\usepackage{epstopdf}
\usepackage{setspace}
\usepackage{subfigure}
\usepackage{natbib}
\usepackage{lscape}
\usepackage{booktabs}

\graphicspath{{./Figs/}}
\usepackage{upgreek}

\usepackage{lineno,hyperref}
\modulolinenumbers[1]
\usepackage{nomencl} 
\usepackage{longtable}
\usepackage{breqn}

\journal{arXiv}

\begin{document}


\begin{frontmatter}



\title{A finite-element model for computing fluid flow inside a sessile evaporating droplet on a solid surface}


\author{Manish Kumar} 
\author{Rajneesh Bhardwaj \corref{mycorrespondingauthor}}
\cortext[mycorrespondingauthor]{Corresponding author}
\ead{rajneesh.bhardwaj@iitb.ac.in}

\address{Department of Mechanical Engineering, Indian Institute of Technology Bombay, Mumbai, 400076, India}

\begin{abstract}
A finite element model was developed to compute the fluid flow inside a sessile evaporating droplet on hydrophilic substrate in ambient conditions. The evaporation is assumed as quasi-steady and the flow is considered as axisymmetric with a pinned contact line. The Navier-Stokes equations in cylindrical coordinates were solved inside the droplet. Galerkin weight residual approach and velocity pressure formulation was used to discretise the governing equations. Six node triangular mesh and quadratic shape functions were used to obtain higher accuracy solutions. Radial velocity profiles in axial directions calculated by the FEM solver were compared with a existing analytical model and were found in excellent agreement. The contours of velocity magnitude and streamlines show the characteristic flow i.e. radially outward inside the evaporating droplet.
\end{abstract}


\end{frontmatter}

\section{Introduction}
\label{sec:intro}
In recent times, investigation of sessile droplet evaporation has been an interesting area of research owing to its many industrial applications such as inkjet printing, spray cooling, making bioassay etc. To understand heat flow inside an evaporating droplet of pure liquid or predict particle deposition pattern in evaporating colloidal droplet, flow inside the droplet needed to be studied and calculation of flow field becomes the necessity. There are two modes of droplet evaporation; (1) constant contact radius (CCR) mode, in which droplet contact line remains pinned during the evaporation, (2) constant contact angle (CCA) mode, in which contact angle of the droplet remains constant and contact line recedes.

Ghasemi and Ward \cite{ghasemi2010energy} experimentally investigated the effect of thermal capillary flow and energy transported by it inside droplet during evaporation and reported it a dominating mode of energy transport near the three-phase line of the droplet, but at the apex of the droplet, conduction was the dominating mode of energy transfer. Using COMSOL Multiphysics V4.3a software, Barmi and Meinhart \cite{barmi2014convective} simulated the convective flow inside the pinned evaporating droplet while assuming axisymmetric flow field, temperature, and vapor concentration distribution. They concluded that flow inside the droplet is responsible for particle deposition patterns but it does not influence heat transfer much and conduction remains the major mode of heat transfer.

In the present work, we consider diffusion-limited and quasi-steady droplet evaporation in CCR mode and presents modeling to calculate the flow inside the evaporating droplet. As it reported in previous studies \cite{hu2005analysis,popov2005evaporative}, droplet evaporates unevenly along the liquid-gas interface, minimum at the top and maximum at the edge. For a micro-liter size droplet surface tension force dominates, it tries to maintain its spherical cap shape. So when there is more loss of liquid at the edge compare to top due to evaporation, droplet shape gets changed and in order to retain its spherical shape, fluid from the top rushes to edge side to compensate the loss. This process generates the fluid flow inside the droplet. The flow inside the droplet is also responsible for different types of deposition patterns on complete drying of the droplet \cite{deegan2000pattern,li2015coffee}.

\section{Modeling}
In this section, the Finite Element Modeling (FEM) of continuity and Navier-Stokes equation will be explained to calculate flow field inside the evaporating droplet. As it explained in previous section, droplet evaporates unevenly along the liquid-gas interface, minimum at the top and maximum at the edge. For a micro-liter droplet surface tension force dominates, it tries to maintain its spherical cap shape. So when there is more loss of liquid at the edge compare to top due to evaporation, droplet shape gets changed and in order to retain its spherical shape, fluid from the top rushes to edge side to compensate the loss. This process generates the fluid flow inside the droplet.

\subsection{Governing equation}

When droplet is very small and fluid flow is very slow, in that case, surface tension forces dominates over shear and normal stress in the droplet and retains a spherical cap shape. For the small droplet of size 1 mm and height of 0.4mm and having the flow velocity of the order of 1 $\mu$m/s, the Bond number (accounts for gravitational force versus surface tension force) is 0.04 and capillary number (accounts for viscous stress versus surface tension forces) is the order of $10^{-8}$ \cite{hu2005analysis}. Therefore, the approximation of spherical cap is justified. Reynolds number is also small ($Re=0.003$), so we can neglect inertial terms from the Navier-Stokes equation. As explained in previous section, our problem is axisymmetric, we can write governing equation of flow (continuity and Navier-Stokes equation) in 2-Dimensional ($r$, $z$) cylindrical coordinates as follows:

\begin{equation}
\frac{1}{r} \frac{\partial (ru)}{\partial r} + \frac{\partial v}{\partial z} = 0
\label{femeqn1}
\end{equation}

\begin{equation}
\frac{1}{r} \frac{\partial (r\sigma_{rr})}{\partial r} + \frac{\partial \sigma_{rz}}{\partial z} - \frac{\sigma_{\theta \theta}}{r}= 0
\label{femeqn2}
\end{equation}

\begin{equation}
\frac{1}{r} \frac{\partial (r\sigma_{zr})}{\partial r} + \frac{\partial \sigma_{zz}}{\partial z} = 0
\label{femeqn3}
\end{equation}

where,

\begin{equation}
\sigma_{rr} = 2\mu \frac{\partial u}{\partial r} - P
\label{femeqn4}
\end{equation}

\begin{equation}
\sigma_{zz} = 2\mu \frac{\partial v}{\partial z} - P
\label{femeqn5}
\end{equation}

\begin{equation}
\sigma_{\theta \theta} = 2\mu \frac{u}{r} - P
\label{femeqn6}
\end{equation}

\begin{equation}
\sigma_{rz} = \sigma_{zr} =  \mu \left( \frac{\partial v}{\partial r} + \frac{\partial u}{\partial z} \right)
\label{femeqn7}
\end{equation}

\begin{figure}
\centering
\includegraphics[width=0.8\textwidth]{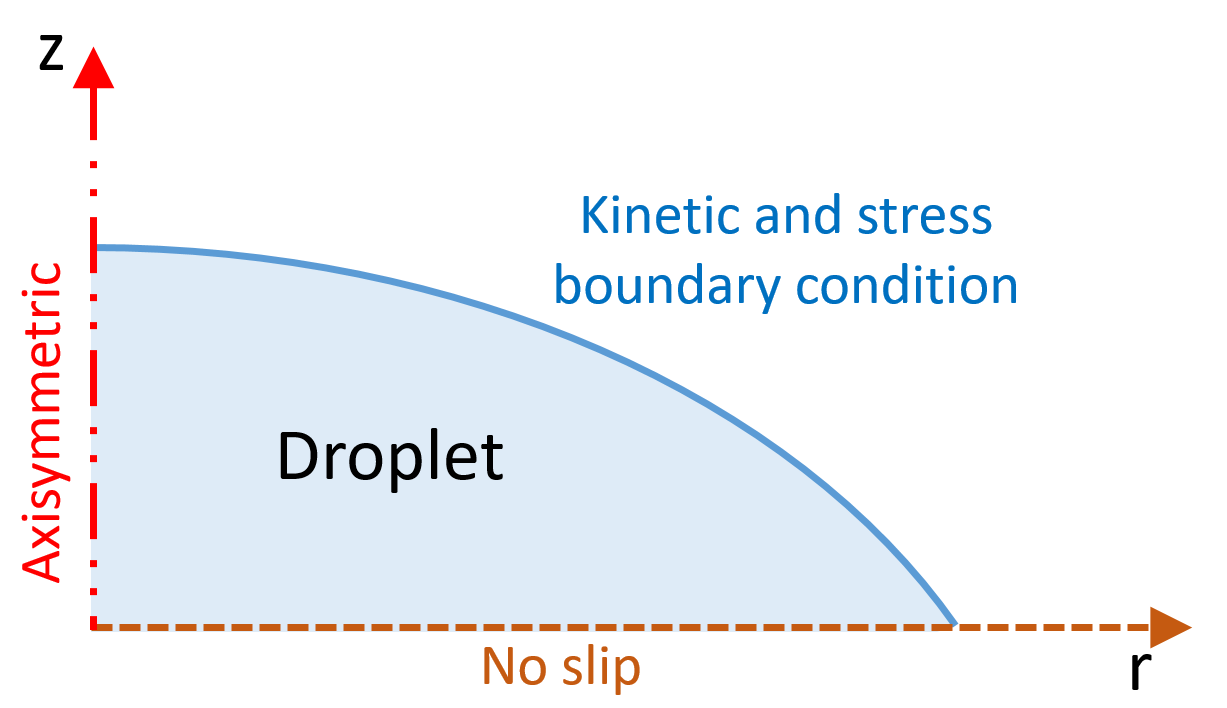}
\caption{Schematic of droplet domain showing boundary conditions at respective boundary.}
\label{domain}
\end{figure}

The boundary conditions (Figure \ref{domain}) for the above equation can be expressed as follows: At the bottom of the droplet (at $z = 0$), no slip boundary condition ($u = 0$, $v = 0$) was applied. At $r = 0$, axisymmetric boundary condition was applied. At the liquid-gas interface, two boundary conditions was applied, shear stress boundary condition in tangential direction and kinematic boundary condition in normal direction.

\subsection{FEM formulation}
In our FEM formulation, we have utilized Galerkin weighted residual approach, in which, weight function ($w_i$) are the same as shape function or approximation function. Using weight function following weak form of Navier-Stokes equation was obtained to lower the requirement of choosing higher order approximation function, which is as follows:

\begin{equation}
\int_{\Omega_e} w_i \left[ \frac{1}{r} \frac{\partial (r\sigma_{rr})}{\partial r} + \frac{\partial \sigma_{rz}}{\partial z} - \frac{\sigma_{\theta \theta}}{r} \right]d\Omega= 0
\label{femeqn8}
\end{equation}

\begin{equation}
\int_{\Omega_e} w_i \left[ \frac{1}{r} \frac{\partial (r\sigma_{zr})}{\partial r} + \frac{\partial \sigma_{zz}}{\partial z} \right]d\Omega= 0
\label{femeqn9}
\end{equation}

where, $d\Omega = 2\pi r dr dz$, differential volume for axisymmetric coordinates and $\Omega_e$ represents a finite element. After solving and rearranging differential term in above integrals, we get the following equations:

\begin{equation}
\int_{\Omega_e} \left[ \frac{\partial w_i (\sigma_{rr})}{\partial r} + \frac{\partial w_i \sigma_{rz}}{\partial z} + \frac{w_i \sigma_{\theta \theta}}{r} \right]rdrdz= \int_{\Gamma_e} w_i (t_r)rds
\label{femeqn10}
\end{equation}

\begin{equation}
\int_{\Omega_e} \left[ \frac{\partial w_i (\sigma_{zr})}{\partial r} + \frac{\partial w_i \sigma_{zz}}{\partial z} \right]rdrdz= \int_{\Gamma_e} w_i (t_z)rds
\label{femeqn11}
\end{equation}

where, $t_r = \sigma_{rr}n_r+\sigma_{rz}n_z$ and $t_z = \sigma_{zr}n_r+\sigma_{zz}n_z$ are the boundary stress component in radial and axial directions, respectively. $\Gamma_e$ represents the boundary of element and differential surface area ($d\Gamma$) of axisymmetric geometry is $2\pi rds$. where, $ds$ is differential arc length of the boundary.

\subsubsection{Penalty function formulation}
In penalty function formulation pressure ($P$) is eliminated from Navier-Stokes equations. The elimination of pressure leads to constrained problem and where constrained equation is the continuity equation (eq. \ref{femeqn1}). Using penalty function method, constrained problem is reformulated as an unconstrained problem and pressure $P$ is replaced by following function \cite{reddyfem}:

\begin{equation}
P = -\gamma \left( \frac{\partial u}{\partial r} + \frac{\partial v}{\partial z} + \frac{u}{r} \right)
\label{femeqn12}
\end{equation}

Where, $\gamma$ is a penalty factor which should be a large arbitrary value \cite{reddyfem}, i.e. $\mu\times 10^{10} $. Note that, After calculating velocities pressure ($P$) can be recovered using above equation (eq. \ref{femeqn12}). Now, on putting the values of $\sigma_{rr}$, $\sigma_{zz}$, $\sigma_{\theta \theta}$ and $P$ from eq. \ref{femeqn4}, \ref{femeqn5}, \ref{femeqn6}, \ref{femeqn7} and \ref{femeqn12} respectively in eq. \ref{femeqn10} and \ref{femeqn11}, following equation can be obtained:

\begin{dmath}
\int_{\Omega_e} \left[ 2\mu \frac{\partial w_i}{\partial r} \frac{\partial u}{\partial r} + \mu \frac{\partial w_i}{\partial z} \left(\frac{\partial u}{\partial z} + \frac{\partial v}{\partial r} \right) + 2\mu \frac{w_i}{r} \frac{u}{r} + 
\gamma \frac{\partial w_i}{\partial r} \left( \frac{\partial u}{\partial r} + \frac{\partial v}{\partial z} + \frac{u}{r} \right) + \gamma \frac{w_i}{r} \left( \frac{\partial u}{\partial r} + \frac{\partial v}{\partial z} + \frac{u}{r} \right) \right]rdrdz= \int_{\Gamma_e} w_i (t_r)rds
\label{femeqn13}
\end{dmath}

\begin{dmath}
\int_{\Omega_e} \left[ \mu \frac{\partial w_i}{\partial r} \frac{\partial u}{\partial z} + \mu \frac{\partial w_i}{\partial r} \frac{\partial v}{\partial r} + 2\mu \frac{\partial w_i}{\partial z} \frac{\partial v}{\partial z} + \gamma \frac{\partial w_i}{\partial z} \left( \frac{\partial u}{\partial r} + \frac{\partial v}{\partial z} + \frac{u}{r} \right) \right]rdrdz= \int_{\Gamma_e} w_i (t_z)rds
\label{femeqn14}
\end{dmath}

To discretize the above continuous equations within each finite element, the velocities ($u$, $v$) are approximated by the following trial solution:

\begin{equation}
\begin{aligned}
u &= \sum_{j=1}^{n} \psi_j(r,z)u_j \\ 
v &= \sum_{j=1}^{n} \psi_j(r,z)v_j
\end{aligned}
\label{femeqn15}
\end{equation}

where $u_j$ and $v_j$ are the radial and axial velocity at the nodal points of the element. $n$ and $\psi_j$ are the number of nodes in the element and shape functions respectively. As explained before, in Galerkin weighted residual approach, weighting functions are same as shape functions, therefore $w_i=\psi_i$. On putting trial solution and weighting function, we can write eq. \ref{femeqn13} and \ref{femeqn14} in compact matrix notation as follows:

\begin{dmath}
\left(2\mu\left[S^{rr}\right]+\mu\left[S^{zz}\right]+2\mu\left[M\right]+\gamma \left(\left[S^{rr}\right]+\left[S^{ro}\right]+\left[S^{or}\right]+\left[M\right]\right)\right)\lbrace u_j \rbrace+
\left(\mu\left[S^{zr}\right]+\gamma\left[S^{rz}\right]+\gamma\left[S^{oz}\right]\right)\lbrace v_j \rbrace = \lbrace F^r \rbrace
\label{femeqn16}
\end{dmath}

\begin{dmath}
\left(\mu\left[S^{rz}\right]+\gamma\left[S^{zr}\right]+\gamma\left[S^{zo}\right]\right)\lbrace u_j \rbrace+ \left(\mu\left[S^{rr}\right]+2\mu\left[S^{zz}\right]+\gamma\left[S^{zz}\right]\right)\lbrace v_j \rbrace = \lbrace F^z \rbrace
\label{femeqn17}
\end{dmath}

Further, we can write above two equations in more simplified way in matrix form as follows:

\begin{dmath}
\begin{bmatrix}
\left[K^{rr}\right] & \left[K^{rz}\right] \\
\left[K^{zr}\right] & \left[K^{zz}\right]
\end{bmatrix}
\begin{Bmatrix}
\lbrace u_j \rbrace\\
\lbrace v_j \rbrace
\end{Bmatrix}
=
\begin{Bmatrix}
\lbrace F^r \rbrace\\
\lbrace F^z \rbrace
\end{Bmatrix}
\label{femeqn18}
\end{dmath}

where,

\begin{equation}
\begin{aligned}
\left[K^{rr}\right] &=  2\mu\left[S^{rr}\right]+\mu\left[S^{zz}\right]+2\mu\left[M\right]+\gamma \left(\left[S^{rr}\right]+\left[S^{ro}\right]+\left[S^{or}\right]+\left[M\right]\right)\\ 
\left[K^{rz}\right] &= \mu\left[S^{zr}\right]+\gamma\left[S^{rz}\right]+\gamma\left[S^{oz}\right]\\
\left[K^{zr}\right] &= \mu\left[S^{rz}\right]+\gamma\left[S^{zr}\right]+\gamma\left[S^{zo}\right]\\
\left[K^{zz}\right] &= \mu\left[S^{rr}\right]+2\mu\left[S^{zz}\right]+\gamma\left[S^{zz}\right]
\end{aligned}
\label{femeqn19}
\end{equation}

where the element coefficient matrices are defined as follows:

\begin{dmath}
M_{ij} = \int_{\Omega_e} \frac{\psi_i \psi_j}{r}drdz
\label{femeqn20}
\end{dmath}

\begin{equation}
\begin{aligned}
S_{ij}^{rr} = \int_{\Omega_e} \frac{\partial\psi_i}{\partial r} \frac{\partial\psi_j}{\partial r}rdrdz 
\hspace{2cm} 
S_{ij}^{rz} = \int_{\Omega_e} \frac{\partial\psi_i}{\partial r} \frac{\partial\psi_j}{\partial z}rdrdz
\end{aligned}
\label{femeqn21}
\end{equation}

\begin{equation}
\begin{aligned}
S_{ij}^{zr} = \int_{\Omega_e} \frac{\partial\psi_i}{\partial z} \frac{\partial\psi_j}{\partial r}rdrdz 
\hspace{2cm} 
S_{ij}^{zz} = \int_{\Omega_e} \frac{\partial\psi_i}{\partial z} \frac{\partial\psi_j}{\partial z}rdrdz
\end{aligned}
\label{femeqn22}
\end{equation}

\begin{equation}
\begin{aligned}
S_{ij}^{ro} = \int_{\Omega_e} \frac{\partial\psi_i}{\partial r} \psi_j drdz 
\hspace{2cm} 
S_{ij}^{or} = \int_{\Omega_e} \psi_i \frac{\partial\psi_j}{\partial r} drdz
\end{aligned}
\label{femeqn23}
\end{equation}

\begin{equation}
\begin{aligned}
S_{ij}^{zo} = \int_{\Omega_e} \frac{\partial\psi_i}{\partial z} \psi_j drdz 
\hspace{2cm} 
S_{ij}^{oz} = \int_{\Omega_e} \psi_i \frac{\partial\psi_j}{\partial z} drdz
\end{aligned}
\label{femeqn24}
\end{equation}

\begin{equation}
\begin{aligned}
F_i^r = \int_{\Gamma_e} psi_i (t_r)rds
\hspace{2cm} 
F_i^z = \int_{\Gamma_e} psi_i (t_z)rds
\end{aligned}
\label{femeqn25}
\end{equation}

\subsubsection{Velocity-Pressure formulation}

Velocity-pressure formulation is a natural and direct formulation \cite{reddyfem}. In this formulation, the weak form of continuity equation is obtained using weighting function, one order less than used for Navier-Stokes equations. Both Navier-Stokes equation physically represents force, hence same weighting function can be used. However, continuity equation represents the volume change. Volume change occur under the action of hydrostatic pressure, hence weight function ($w_{2k}$) for continuity, equation should like the pressure ($P$) or the shape function of pressure. The weak form of the continuity equation can be written as follows:

\begin{dmath}
- \int_{\Omega_e} w_{2k} \left[ \frac{\partial u}{\partial r} + \frac{\partial v}{\partial z} + \frac{u}{r} \right]rdrdz = 0
\label{femeqn26}
\end{dmath}

The minus sign is inserted to make the resulting finite element model symmetric. Now, on putting the values of   from eq. \ref{femeqn4}, \ref{femeqn5}, \ref{femeqn6} and \ref{femeqn7}, respectively in eq. \ref{femeqn10} and \ref{femeqn11}, and expanding eq. \ref{femeqn26}, following equations can be obtained:

\begin{dmath}
\int_{\Omega_e} \left[ 2\mu \frac{\partial w_i}{\partial r} \frac{\partial u}{\partial r} + \mu \frac{\partial w_i}{\partial z} \left(\frac{\partial u}{\partial z} + \frac{\partial v}{\partial r} \right) + 2\mu \frac{w_i}{r} \frac{u}{r} - \frac{\partial w_i}{\partial r}P - \frac{w_i}{r}P \right]rdrdz= \int_{\Gamma_e} w_i (t_r)rds
\label{femeqn27}
\end{dmath}

\begin{dmath}
\int_{\Omega_e} \left[ \mu \frac{\partial w_i}{\partial r} \frac{\partial u}{\partial z} + \mu \frac{\partial w_i}{\partial r} \frac{\partial v}{\partial r} + 2\mu \frac{\partial w_i}{\partial z} \frac{\partial v}{\partial z} - \frac{\partial w_i}{\partial z}P \right]rdrdz= \int_{\Gamma_e} w_i (t_z)rds
\label{femeqn28}
\end{dmath}

\begin{dmath}
- \int_{\Omega_e} \left[ w_{2k} \frac{\partial u}{\partial r} + w_{2k} \frac{\partial v}{\partial z} + w_{2k} \frac{\partial u}{\partial r} \right]rdrdz= 0
\label{femeqn29}
\end{dmath}

To discretize the above continuous equations within each finite element, the velocities ($u$, $v$) are approximated by the trial solution defined in eq. \ref{femeqn15} and pressure ($P$) is approximated by following trial solution, which one order less than used for velocities:

\begin{equation}
P = \sum_{l=1}^{m} \phi_l(r,z)P_l
\label{femeqn30}
\end{equation}

where $P_l$ is the pressure at the corner nodal points of the element. $m$ and $\phi_l$ are the number of corner nodes in the element and shape functions respectively. As explained before, in Galerkin weighted residual approach, $w_i=\psi_i$ and $w_{2k}= \phi_l$. On putting trial solution and weighting function, we can write eq. \ref{femeqn27}, \ref{femeqn28} and \ref{femeqn29} in compact matrix notation as follows:

\begin{dmath}
\left(2\mu\left[S^{rr}\right]+\mu\left[S^{zz}\right]+2\mu\left[M\right] \right)\lbrace u_j \rbrace + \left(\mu\left[S^{zr}\right]\right)\lbrace v_j \rbrace - \left(\left[G^{ro}\right]+\left[G^{oo}\right]\right)\lbrace P_l \rbrace= \lbrace F^r \rbrace
\label{femeqn31}
\end{dmath}

\begin{dmath}
\left(\mu\left[S^{rz}\right]\right)\lbrace u_j \rbrace+ \left(\mu\left[S^{rr}\right]+2\mu\left[S^{zz}\right]\right)\lbrace v_j \rbrace - \left(\left[G^{zo}\right]\right)\lbrace P_l \rbrace= \lbrace F^z \rbrace
\label{femeqn32}
\end{dmath}

\begin{dmath}
- \left(\left[G^{ro}\right]^T+\left[G^{oo}\right]^T\right)\lbrace u_j \rbrace - \left(\left[G^{zo}\right]^T\right)\lbrace v_j \rbrace= 0
\label{femeqn33}
\end{dmath}

Further, we can write above two equations in more simplified way in matrix form as follows:

\begin{dmath}
\begin{bmatrix}
\left[K^{rr}\right] & \left[K^{rz}\right] & \left[K^{ro}\right]\\
\left[K^{zr}\right] & \left[K^{zz}\right] & \left[K^{zo}\right]\\
\left[K^{or}\right] & \left[K^{oz}\right] & \left[0\right]
\end{bmatrix}
\begin{Bmatrix}
\lbrace u \rbrace\\
\lbrace v \rbrace\\
\lbrace P \rbrace
\end{Bmatrix}
=
\begin{Bmatrix}
\lbrace F^r \rbrace\\
\lbrace F^z \rbrace\\
\lbrace 0 \rbrace
\end{Bmatrix}
\label{femeqn34}
\end{dmath}

where,

\begin{equation}
\begin{aligned}
\left[K^{rr}\right] &=  2\mu\left[S^{rr}\right]+\mu\left[S^{zz}\right]+2\mu\left[M\right]\\ 
\left[K^{rz}\right] &= \mu\left[S^{zr}\right]\\
\left[K^{ro}\right] &= - \left(\left[G^{ro}\right]+\left[G^{oo}\right]\right)\\
\left[K^{zr}\right] &= \mu\left[S^{rz}\right]\\
\left[K^{zz}\right] &= \mu\left[S^{rr}\right]+2\mu\left[S^{zz}\right]\\
\left[K^{zo}\right] &= -\left[G^{zo}\right]\\
\left[K^{or}\right] &=  - \left(\left[G^{ro}\right]^T+\left[G^{oo}\right]^T\right)\\
\left[K^{oz}\right] &= -\left[G^{zo}\right]^T
\end{aligned}
\label{femeqn35}
\end{equation}

where the $[S]$, $[M]$ and $\lbrace F \rbrace$ element coefficient matrices are defined as before from eq. \ref{femeqn20} to eq. \ref{femeqn25}. The remaining element coefficient matrices are defined as follows:

\begin{dmath}
G_{kl}^{ro} = \int_{\Omega_e} \frac{\partial\psi_k}{\partial r} \phi_l  rdrdz
\label{femeqn36}
\end{dmath}

\begin{dmath}
G_{kl}^{zo} = \int_{\Omega_e} \frac{\partial\psi_k}{\partial z} \phi_l  rdrdz
\label{femeqn37}
\end{dmath}

\begin{dmath}
G_{kl}^{oo} = \int_{\Omega_e} \psi_k \phi_l drdz
\label{femeqn38}
\end{dmath}

\subsubsection{Evaluation of element coefficient matrices}

\begin{figure}
\centering
\includegraphics[width=0.7\textwidth]{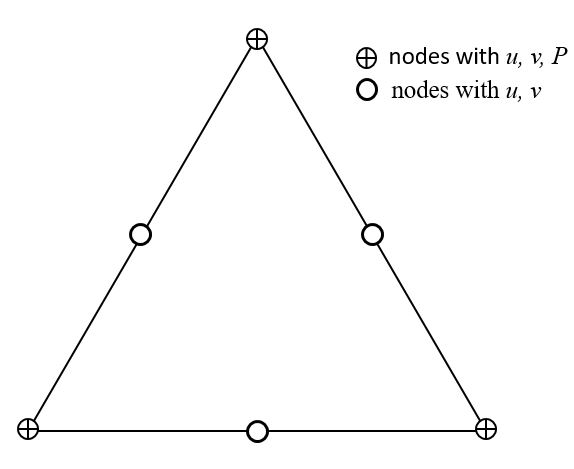}
\caption{Six-node triangular element.}
\label{node6}
\end{figure}

In this model, a six-node triangular element (Figure \ref{node6}) was chosen over three-node element as it improves the solution accuracy and reduce the overall solution time by reducing the number element required. Quadratic shape functions were used to evaluate velocity field and liner shape functions were used to evaluate pressure field. Six-node triangular element contains a node on the mid-side of each of the three side of the triangle, which defines the boundary of the element.

In order to evaluate the integrals of element coefficient matrices, numerical integration was employed. To do numerical integration, first we need to map real element in ($r$, $z$) coordinates to a master or parent element in generic ($\xi$, $\eta$) coordinates over which integration will performed using Gaussian quadrature points. As in our real element all three nodes located at side of triangle are located at mid-point of each triangle side, so it is sufficient to use linear shape function to do coordinate transformation. The following expression facilitates the ($r$, $z$) to ($\xi$, $\eta$)coordinate transformation:

\begin{equation}
\begin{aligned}
r = \sum_{k=1}^{3} r_k \phi_k (\xi, \eta)
\hspace{2cm}
z = \sum_{k=1}^{3} z_k \phi_k (\xi, \eta)
\end{aligned}
\label{femeqn39}
\end{equation}

where $r_k$ and $z_k$ is $k_{th}$ radial and axial coordinates of corner nodes of the triangle element. Shape functions $\phi_k$ can be defined as follows:

\begin{equation}
\begin{aligned}
\phi_1 &= \xi \\
\phi_2 &= \eta \\
\phi_3 &= 1-\eta-\xi
\end{aligned}
\label{femeqn40}
\end{equation}

To evaluate integrals, the following expression is required to map infinitesimal area of real element to corresponding area of parent element:

\begin{equation}
drdz = det\left[J\right] d\xi d\eta
\label{femeqn41}
\end{equation}

where, $det[J]$ is the determinant of the Jacobian ($[J]$) for the element in consideration and is defined as follows:

\begin{dmath}
\left[J\right] = 
\begin{bmatrix}
\frac{\partial r}{\partial \xi} & \frac{\partial z}{\partial \xi}\\
\frac{\partial r}{\partial \eta} & \frac{\partial z}{\partial \eta}
\end{bmatrix}
\label{femeqn42}
\end{dmath}

For linear mapping $det[J]$ is equal to 2 times area of triangle in consideration. Next, we need to convert partial derivative of ($r$, $z$) into derivatives of ($\xi$,$\eta$) which can be done using following relationship:

\begin{dmath}
\begin{Bmatrix}
\frac{\partial \psi_i}{\partial r}\\
\frac{\partial \psi_i}{\partial z}
\end{Bmatrix}
=\left[J\right]^{-1}
\begin{Bmatrix}
\frac{\partial \psi_i}{\partial \xi}\\
\frac{\partial \psi_i}{\partial \eta}
\end{Bmatrix}
\label{femeqn43}
\end{dmath}

where, shape function $\psi_i$ is quadratic shape function for a six-node triangle element as mentioned before. Quadratic shape can be defined as follows:

\begin{equation}
\begin{aligned}
\psi_1 &= 2\xi^2-\xi \\
\psi_2 &= 2\eta^2-\eta \\
\psi_3 &= 1-3\xi-3\eta+4\xi\eta+2\xi^2+2\eta^2\\
\psi_4 &= 4\xi\eta\\
\psi_5 &= 4(\eta-\xi\eta-\eta^2)\\
\psi_6 &= 4(\xi-\xi\eta-\xi^2)
\end{aligned}
\label{femeqn44}
\end{equation}

Using eq. \ref{femeqn41}, \ref{femeqn43} and \ref{femeqn44}, the integrals specified in equations from eq. \ref{femeqn20} to \ref{femeqn24} can be converted to ($\xi$,$\eta$) coordinates and numerical integration can be performed using Gaussian quadrature as follows:

\begin{equation}
\int_{\Omega_e}f(\xi, \eta) d\xi d\eta \cong \frac{1}{2} \sum_{l=1}^{gp} w_l f(\xi_l, \eta_l)
\label{femeqn45}
\end{equation}

where $f(\xi, \eta)$ is any function in ($\xi$, $\eta$) coordinates and ($\xi_l$, $\eta_l$) are the Gaussian points which are well document in any FEM textbook as given in \cite{reddyfem}. $gp$ is the number of Gaussian points used to evaluate the integral. Three and one-point Gaussian quadrature was used to evaluate integrals.

To calculate coefficient of force vector on the right hand side of eq. \ref{femeqn18} or to apply stress boundary condition on the boundary of the domain, we need to transform the boundary or open side of the triangle to one dimensional line coordinate ($s$) whose origin at first point of boundary. As we are using six-node or quadratic triangular element, we have three nodes on each side of the triangle, so we need to use quadratic shape function ($\psi_{bi}$). These three shape function ($\psi_{bi}$) can be defined as follows:

\begin{equation}
\begin{aligned}
\psi_{b1} &= \left(1-\frac{s}{h}\right)\left(1-\frac{2s}{h}\right) \\
\psi_{b2} &= \frac{4s}{h}\left(1-\frac{s}{h}\right) \\
\psi_{b3} &= -\frac{s}{h}\left(1-\frac{2s}{h}\right)
\end{aligned}
\label{femeqn46}
\end{equation}

where, $h$ is the length of the side or boundary of the element. The elements which are inside the domain, for them, the value of coefficients in force vector are need not to be calculated as the force acting on the side or boundary of the neighboring elements cancels out each other. Therefore, we need to calculate coefficient of force vector only at the boundary of elements that are located at boundary of the domain and some external force or stress is applied.

\subsubsection{Boundary conditions}

At the bottom of the droplet (at $z = 0$), no slip boundary condition ($u = 0, v = 0$) was applied, which is a Dirichlet boundary condition. At $r = 0$, axisymmetric boundary condition was applied, which is $u = 0$ and zero stress or velocity gradient. At the liquid-gas interface, two boundary conditions were applied, shear stress boundary condition in tangential direction ($\tau_t$) and kinematic boundary condition in normal direction ($u_n$). The liquid-gas interface boundary condition can be applied using two different approach and both were tested. These approaches are described as follows:\\

\noindent{\it{\textbf{Direct approach}}}

As the shape of droplet is spherical cap, the liquid-gas interface is not parallel to $r$ or $z$ axis. To apply the both boundary conditions precisely, coordinate rotation must be done for the nodes at the liquid-gas interface of the droplet \cite{reddyfem}, which can be done with the help of rotation matrix $[Q]$. Eq. \ref{femeqn18} can be written in more compact form as follows:

\begin{equation}
\left[K\right] \lbrace u \rbrace = \lbrace F \rbrace
\label{femeqn47}
\end{equation}

To perform the coordinate rotation on the nodes at the liquid-gas interface of the droplet following mathematical operation can be done:

\begin{equation}
\left[Q\right]^T\left[K\right]\left[Q\right] \lbrace u \rbrace = \lbrace F \rbrace
\label{femeqn48}
\end{equation}

where, rotation matrix [Q] can be defined as follows:

\begin{dmath}
\left[Q\right]
=
\begin{bmatrix}
\ddots & \cdots & \cdots & \cdots\\
\cdots & n_{r,i} & n_{z,i} & \cdots\\
\cdots & -n_{z,i} & n_{r,i} & \cdots\\
\cdots & \cdots & \cdots & \ddots\\
\end{bmatrix}
\label{femeqn49}
\end{dmath}

where, $n_{r,i}$ and $n_{z,i}$ are the normal vector in radial and axial direction for $i_{th}$ node on the liquid-gas interface. $n_{r}$ and $n_{z}$ can be calculated as follows:

\begin{equation}
\begin{aligned}
n_{r} &= \frac{r\sin\theta}{R} \\
n_{z} &= \sqrt{1-(n_r)^2}
\end{aligned}
\label{femeqn50}
\end{equation}

where $r$ is the radial distance of the considered node and $\theta$ is the contact angle of the droplet. In the normal direction of the liquid-gas interface, kinematic boundary condition can be defined as follows \cite{hu2005analysis}:

\begin{equation}
u_n = \frac{j}{\rho}-n_z\frac{\partial h}{\partial t}
\label{femeqn51}
\end{equation}

where $u_n$ is the normal velocity at the liquid-gas interface, $j/\rho$ is outflow velocity of droplet evaporation and $\partial h/ \partial t$ is the surface motion of spherical cap to retain its spherical shape. $\partial h/ \partial t$ can be defined as follows \cite{hu2005analysis}:

\begin{equation}
\frac{\partial h }{\partial t }= \dot{h_d} \left[\frac{h_d^4-R^4}{4h_d^3\sqrt{\left(\frac{h_d^2+R^2}{2h_d}\right)^2-r^2}} + \frac{R^2-h_d^2}{2h_d^2}\right]
\label{femeqn52}
\end{equation}

where $\dot{h_d}$ is the velocity of the top of the evaporating droplet, which can be calculated as follows \cite{barmi2014convective}

\begin{equation}
\dot{h_d} = \frac{2\dot{m}}{\rho\pi(R^2+h_d^2)}
\label{femeqn53}
\end{equation}

In the tangential direction, we apply zero shear stress boundary condition (i.e. $\tau_t= t_z = 0$).\\

\noindent{\it{\textbf{Penalty approach}}}

In this approach, kinetic boundary condition is treated as constrained equation, which can be written as follows:

\begin{equation}
u_r n_r + u_z n_z = u_n
\label{femeqn54}
\end{equation}

where $u_r$ and $u_z$ are the velocity component in $r$ and $z$ directing at the liquid-gas interface. Using penalty function method governing equations were solved and the constraint equation (eq. \ref{femeqn54}) is made to satisfy, which yields solution in following form \cite{reddyfem}:

\begin{equation}
\left(\left[K\right] + \left[K_P\right]\right) \lbrace u \rbrace = \lbrace F \rbrace + \lbrace F_P \rbrace
\label{femeqn55}
\end{equation}

where,

\begin{dmath}
\left[K_P\right]
=
\begin{bmatrix}
\cdots & \cdots & \cdots & \cdots\\
\cdots & \gamma_p n_r^2 & \gamma_p n_r n_z & \cdots\\
\cdots &\gamma_p n_r n_z & \gamma_p n_z^2 & \cdots\\
\cdots & \cdots & \cdots & \cdots\\
\end{bmatrix}
\label{femeqn56}
\end{dmath}

\begin{dmath}
\lbrace F_P \rbrace
=
\begin{Bmatrix}
\cdots\\
\gamma_p u_n n_r \\
\gamma_p u_n n_z \\
\cdots 
\end{Bmatrix}
\label{femeqn57}
\end{dmath}

Where, $\gamma$ is a penalty factor which should be a large arbitrary value \cite{reddyfem}, i.e. $max([K])\times 10^{6} $. Thus, a modification of coefficient in $[K]$ and $\lbrace F \rbrace$ matrix associated with boundary nodes will yield the desired solution with the constraint kinetic boundary condition.

Zero shear stress boundary condition can be applied by dividing the tangential shear stress into $r$ and $z$ components, i.e. $t_r = \tau_t n_r = 0$, $t_z = \tau_t n_z = 0$.

\begin{figure}
\centering
\includegraphics[width=1\textwidth]{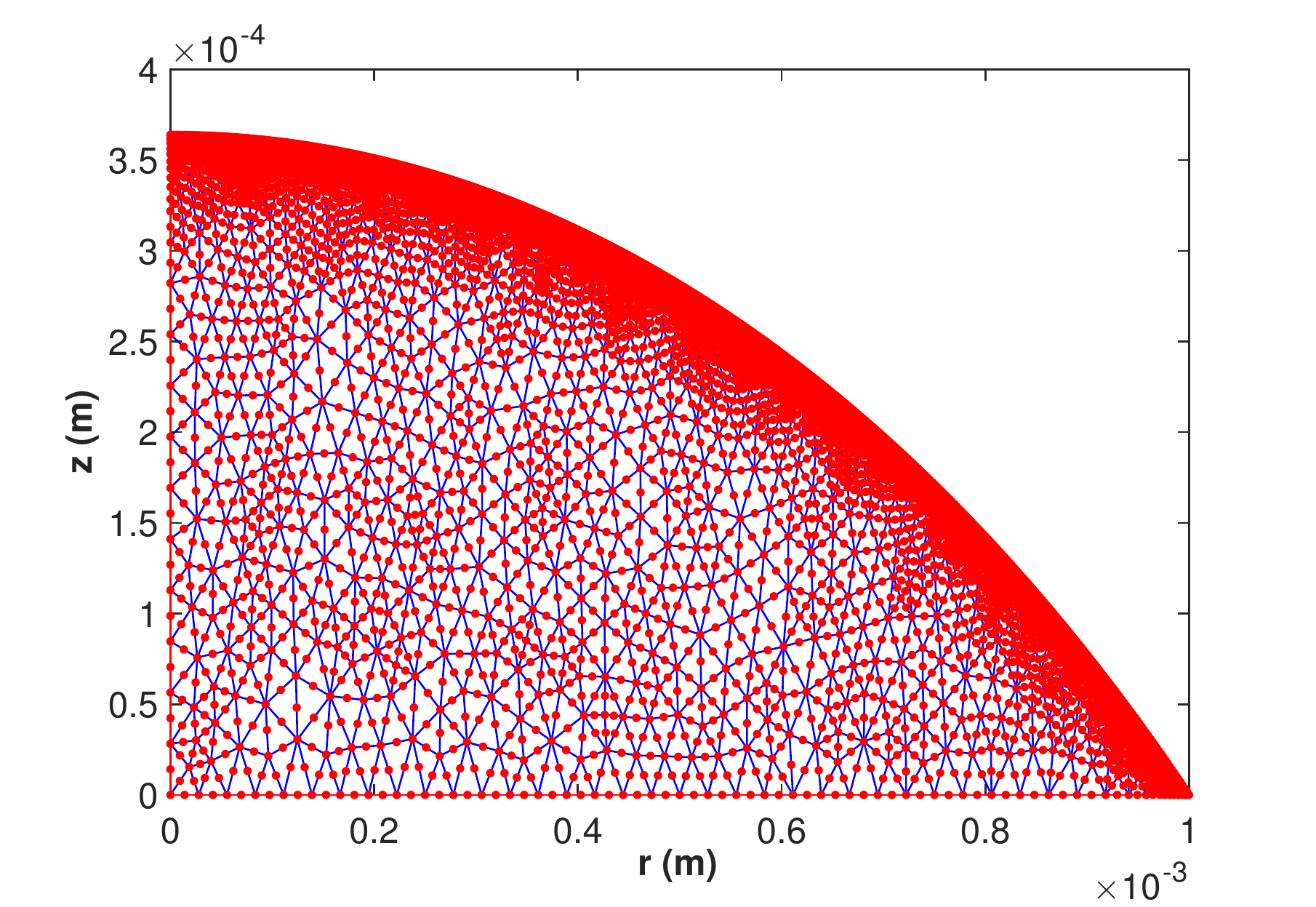}
\caption{Six-node mesh in droplet domain. Red dots represents nodes.}
\label{mesh6node}
\end{figure}

\subsubsection{Mesh generation}
As the whole code in written in MATLAB, we were more inclined to use MATLAB functions to save ourselves from any compatibility and data import issues. To generate six-node triangular element mesh, first we generated a three-node triangular mesh inside the droplet domain using MATLAB function ‘initmesh’. To the best of author knowledge, there is no MATLAB function available to generate six-node triangular mesh. Therefore, to generate six-node mesh, we developed an in-house code, which places a node at the exact midpoint of the each side of triangular element and updates the point matrix, element connectivity matrix, boundary matrix. The final mesh generated using in-house code and MATLAB function `initmesh’ is shown in Figure \ref{mesh6node}.

\section{Results and Validation of code}

We have used two formulations (Penalty function formulation and Velocity-pressure formulation) and two approaches (Direct and Penalty approach) of boundary condition application to solve flow filed inside the droplet, which makes four combinations of methods to solve flow filed. Here, we are presenting validation using direct approach of boundary condition application for both formulations to avoid repeatability of the similar results of penalty approach application of boundary condition.

\begin{figure}
\centering
\includegraphics[width=0.8\textwidth]{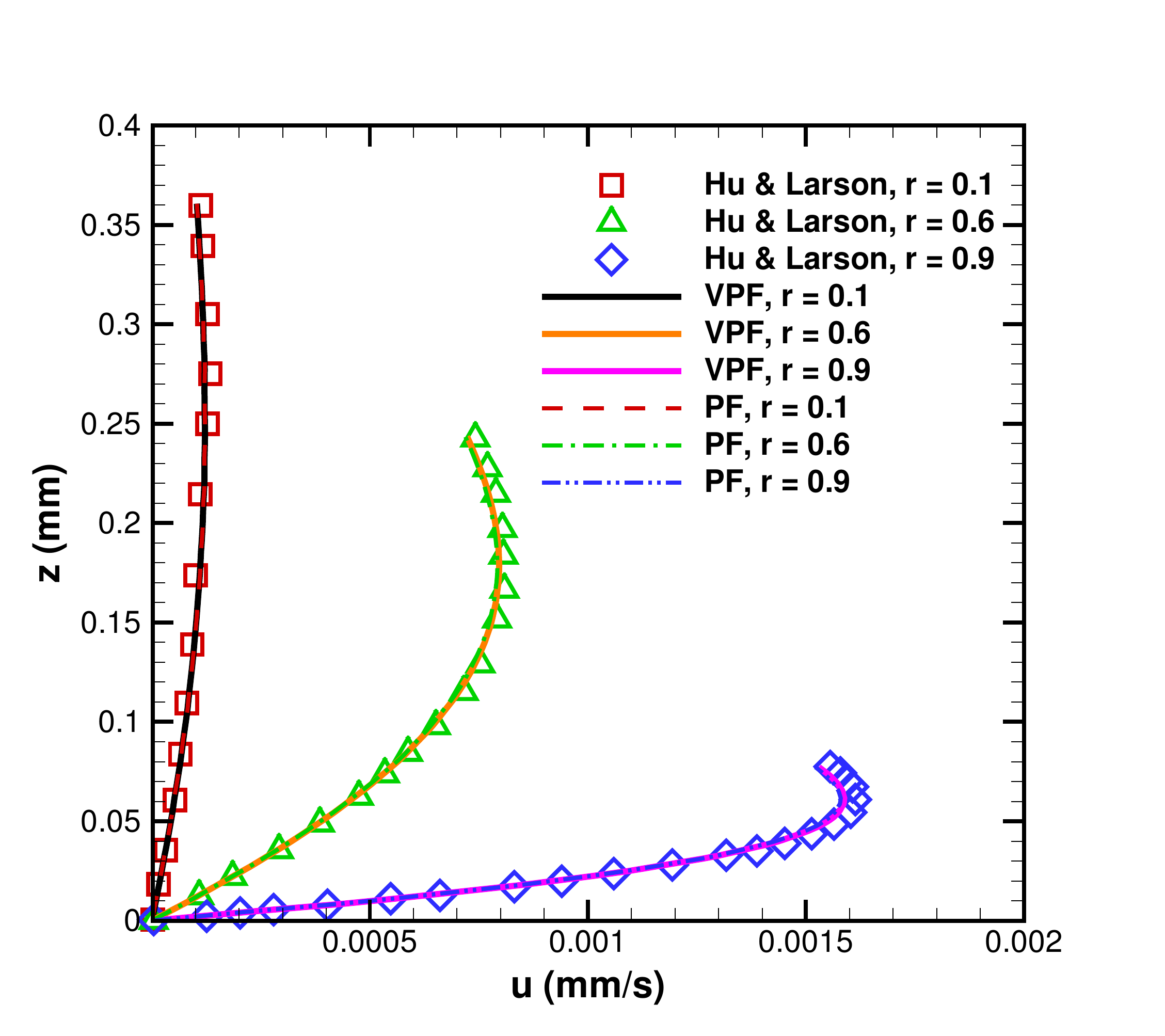}
\caption{Validation of Velocity-pressure and Penalty function formulation: comparison of radial velocity profiles (at different radial locations) obtained from present model with the model of Hu and Larson \cite{hu2005analysis}.}
\label{comp_pd}
\end{figure}




\subsection{Validation of different formulations}
To validate the code radial velocity at different radial location along the height of the droplet has been compared with results of model of Hu and Larson \cite{hu2005analysis}. Wetting angle and wetting radius of droplet considered are $40^\circ$ and 1 mm. Figure \ref{comp_pd} shows the radial velocity ($u$) profiles calculated along axial direction ($z$) at different radial locations, $r$ = 0.1, 0.6 and 0.9 mm, using Velocity-pressure and Penalty function formulation with direct approach of applying boundary conditions. These profiles were compared with the profiles obtained by Hu and Larson \cite{hu2005analysis}. The radial velocity profile obtained from present models are in good agreement with the profiles obtained by Hu and Larson \cite{hu2005analysis}. 


\begin{figure}
\centering
\includegraphics[width=0.8\textwidth]{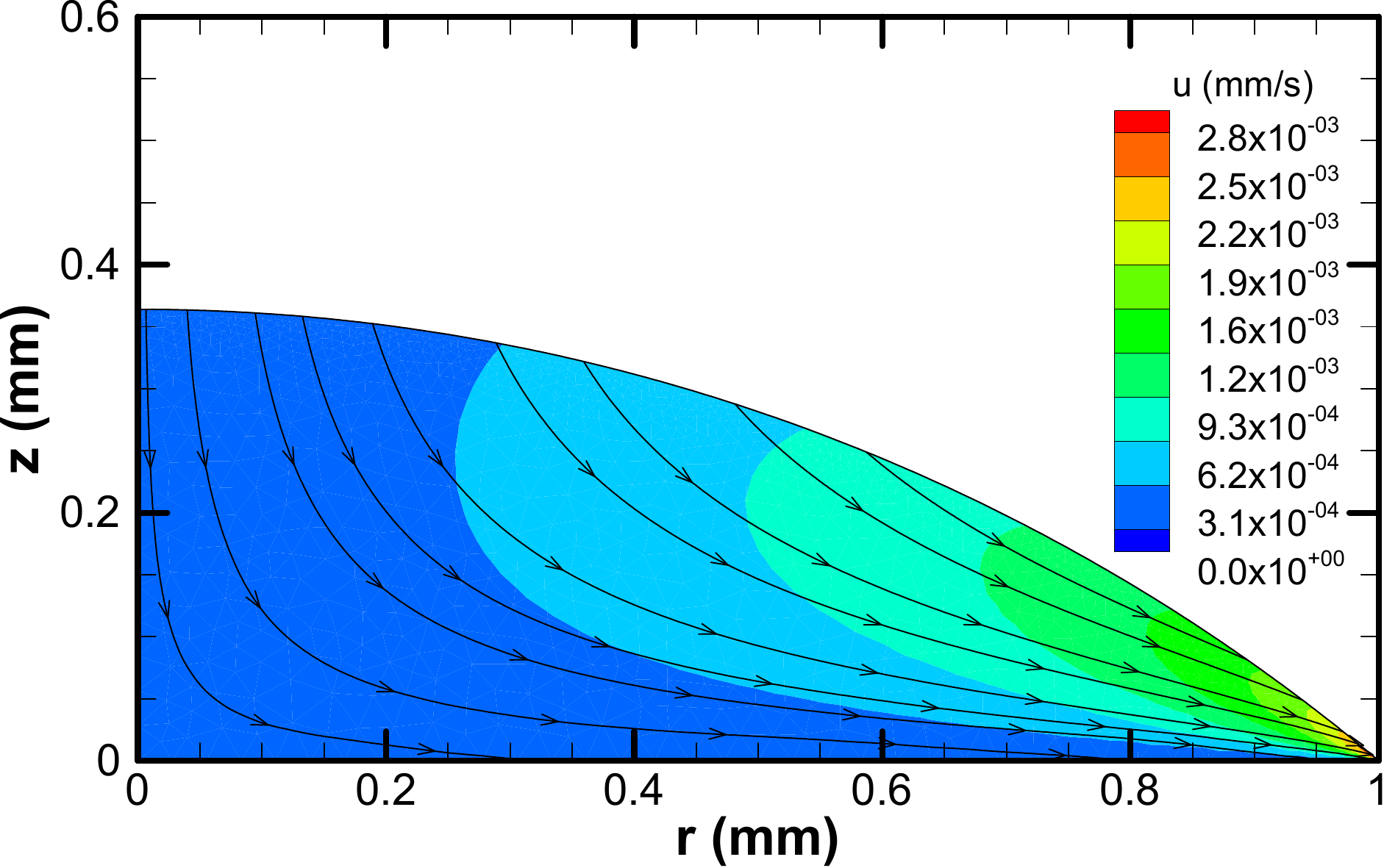}
\caption{Streamlines and $u$-radial velocity contour inside the evaporating droplet.}
\label{streamline}
\end{figure}

\begin{figure}
\centering
\includegraphics[width=0.8\textwidth]{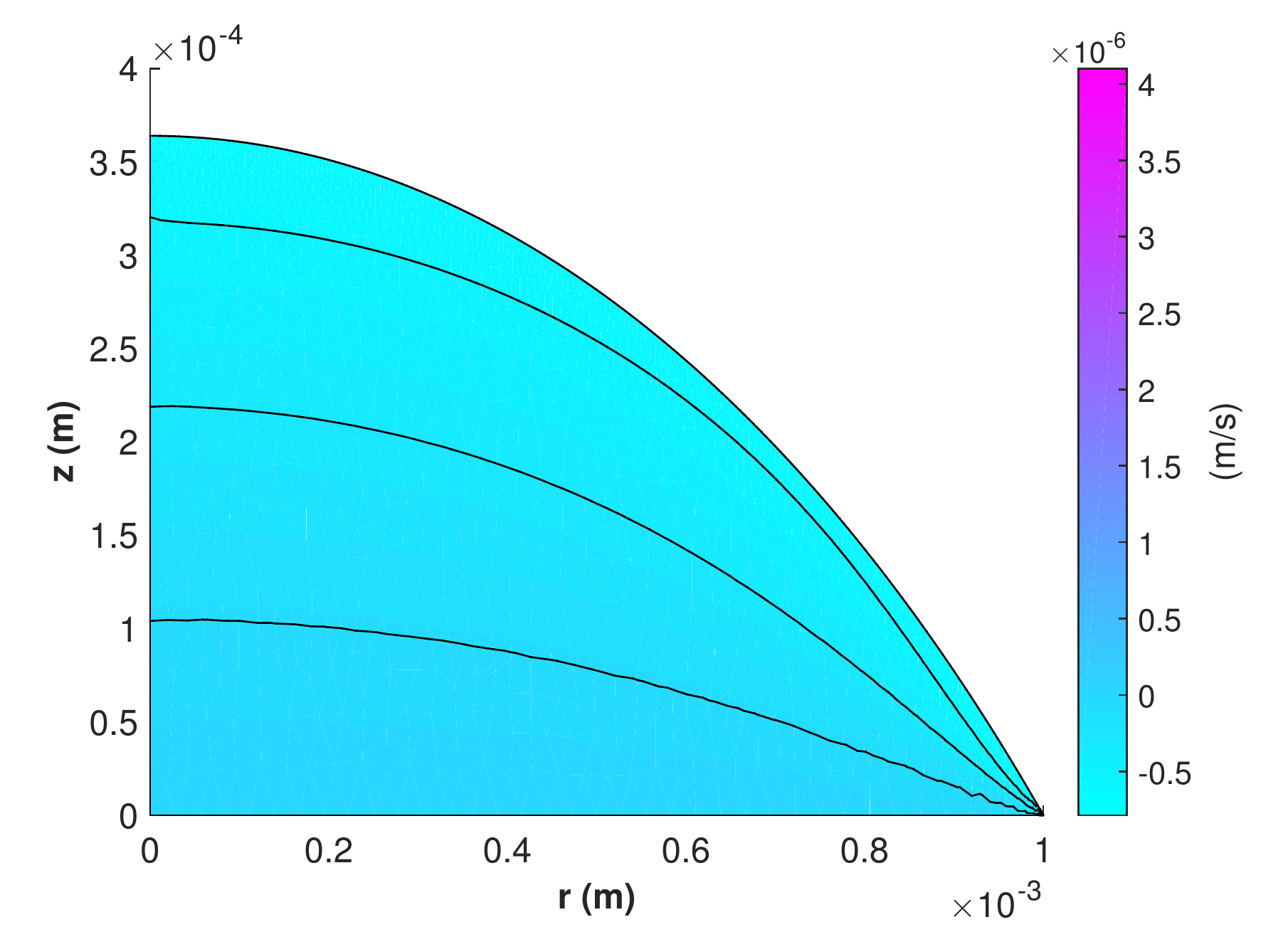}
\caption{Axial velocity ($v$) field inside the evaporating droplet.}
\label{axial}
\end{figure}

\begin{figure}
\centering
\includegraphics[width=0.8\textwidth]{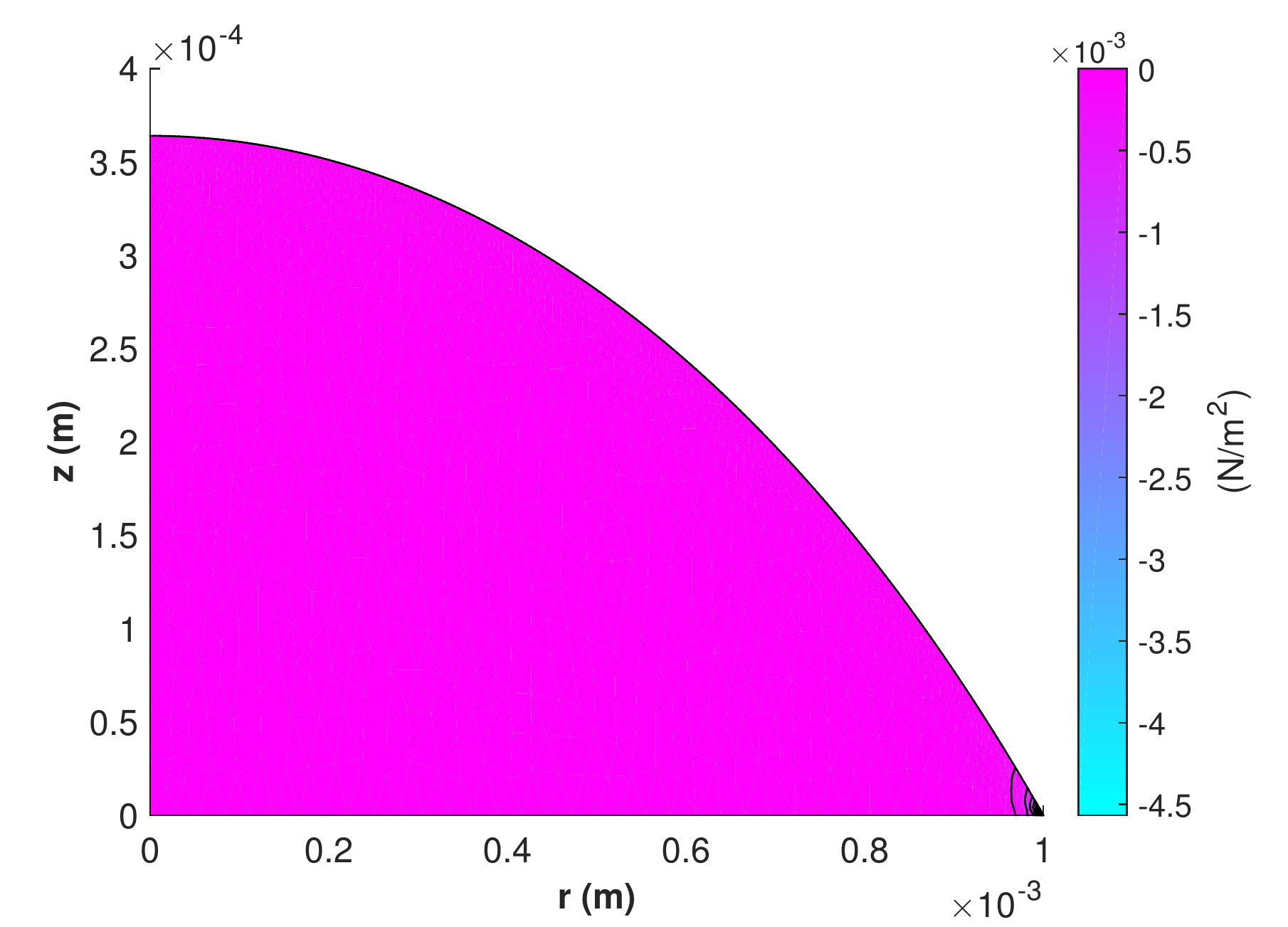}
\caption{Pressure field ($P$) inside the evaporating droplet.}
\label{press}
\end{figure}

\subsection{Flow field inside the droplet}

Fluid velocity inside the droplet increases as flow moves from droplet interior to the three-phase contact line. This phenomenon is evident from three velocity profiles shown in Figure \ref{comp_pd}. At radial location $r$ = 0.1 mm, maximum velocity is $0.13 \times 10^{-3}$ mm/s and as fluid move radially outward at location $r$ = 0.6 and 0.9 mm radial velocity of fluid increases to $0.8 \times 10^{-3}$ mm/ s and $1.6 \times 10^{-3}$ mm/s, respectively. However, in axial direction, from $z$ = 0 to liquid-air interface of droplet, first, droplet velocity increases with the increase of axial distance and after attaining a maximum value it starts decreasing.

Figure \ref{streamline} shows the streamlines and contours of radial velocity. All streamlines are going radially outward towards the three-phase contact line of the droplet and radial velocity increases in radially outward direction and maximum near the contact line. At the contact line evaporation rate is higher compare to remaining liquid-air interface of the droplet \cite{hu2002evaporation},  which results in higher mass loss compare at three-phase contact line compare to remaining interface. As the droplet size falls in capillary regime, droplet tries to maintain its spherical cap shape. Therefore, fluid from the interior of the droplet rushes towards three-phase contact line to maintain droplet spherical cap shape. Streamline shown in Figure \ref{streamline} shows the same behaviour and characteristics and corroborate the above explanation regarding fluid flow inside the droplet.

Figure \ref{axial} shows the axial velocity ($v$) contour inside the evaporating droplet. It can be concluded from the figure that magnitude of $v$ increases away from the droplet-air interface to droplet-substrate contact. Figure \ref{press} shows the pressure ($P$) contour inside the droplet. it can be observed that pressure inside the droplet doesn't vary much, it remains almost same except at the contact line.

\section{Conclusions}
Present work explains the finite element modeling of fluid flow inside evaporating sessile droplet resting on hydrophilic substrate. In this modeling, constant contact radius (CCR) mode of evaporation was considered. Finite element code was implement in MATLAB. Galerkin weighted residual approach was used to formulate weak form of the numerical equations. Velocity-pressure formulation was applied to discretise the Navier-Stokes equation in cylindrical coordinate system. Six-node triangular mesh was used in the simulations for higher accuracy of the solution. Results obtained from the simulations were compared with the model of Hu and Larson \cite{hu2005analysis} and good agreement was found in present work and the model of Hu and Larson.

%
%

\bibliography{ManishBib.bib}

\end{document}